\documentstyle[seceq]{ptptex}

\makeatletter
\def\vereq#1#2{\lower3pt\vbox{\baselineskip1.5pt \lineskip1.5pt
\ialign{$\m@th#1\hfill##\hfil$\crcr#2\crcr\sim\crcr}}}
\notypesetlogo  

\notypesetlogo  

\title{
Grand Unified Models on $S_1/(Z_2 \times Z_2')$ 
 Orbifold\footnote{This papaer is based on the work 
 collaborated with T. Kondo, Y. Shimizu, T. Suzuki, and 
 K. Ukai. }
}

\author{
Naoyuki {\sc Haba}
}

\inst{
{}Faculty of Engineering, Mie University,
Tsu Mie 514-8507, Japan
}

\abst{
\noindent
We suggest two grand unified models 
 where the fifth dimensional coordinate is 
 compactified on an $S_1/(Z_2 \times Z_2')$ 
 orbifold.

}

\begin{document}

\maketitle

\noindent
We suggest two grand unified models 
 where the fifth dimensional coordinate is 
 compactified on an $S_1/(Z_2 \times Z_2')$ orbifold. 
Before showing the models, let us see the compactification
 of the fifth dimension. 
We denote the five dimensional coordinate 
 as $y$, which is 
 compactified on an $S_1/(Z_2 \times Z_2')$ 
 orbifold. 
Under the parity transformation of $Z_2$  and $Z_2'$, 
 which transform $y \rightarrow -y$ and 
 $y' \rightarrow -y'$ ($y'=y+ \pi R/2$), respectively, 
 a field $\phi(x^\mu,y)$ 
transforms as 
$\phi(x^\mu,y) \to \phi(x^\mu,-y) = P\phi(x^\mu,y)$ 
and
$\phi(x^\mu,y') \to \phi(x^\mu,-y') = P'\phi(x^\mu,y')$, 
where $P$ and $P'$ are operators of 
 $Z_2$  and $Z_2'$ transformations, respectively. 
Two walls at $y=0$ $( \pi R)$ and $\pi R/2$ $(- \pi R/2)$ are
 fixed points under $Z_2$ and $Z_2'$ transformations, 
 respectively. 
The physical space can be taken to be $0 \leq y \leq \pi R/2$,  
 since the walls at $y = \pi R$ and $- \pi R/2$ are 
 identified with those at $y = 0$ and $\pi R/2$, respectively. 
On this orbifold, the field $\phi(x^\mu,y)$ 
 is divided into four eigenfunctions 
 according to the eigenvalues $( \pm, \pm)$ 
 of the parity 
 $(Z_2, Z_2')$. 
We take the $Z_2$ parity operator as $P= diag.(1,1,1,1,1)$ and 
 the $Z_2'$ parity operator as $P'= diag.(-1,-1,-1,1,1)$ 
 acting on a ${\bf 5}$ representation in 
 $SU(5)$\cite{kawamura}\cite{kawamura2}\cite{HN}\cite{others}. \\

\noindent
{\bf (1) non-SUSY $SU(5)$ GUT\cite{HKSSU}:} 

The first model we propose is 
 the non-supersymmetric model with the 
 gauge coupling unification. 
We assume that 
 chiral matter fields are localized on the 
 four dimensional wall at $y=0$ $(\pi R)$. 
 Higgs and gauge fields can propagate 
 in five dimensions. 
When we consider the non-supersymmetric 
 $SU(5)$ grand unified theory (GUT), 
 one of the most serious problems is 
 that the gauge coupling constants are not
 unified at the high energy. 
However, the gauge coupling unification can be
 realized by a simple extension in the 
 standard model proposed in Ref.\cite{MY}, 
 where one more Higgs doublet and 
 two leptoquark scalar fields with the quantum numbers of 
 $({\bf 3}, {\bf 2})+({\bf \overline{3}}, {\bf 2})$ in 
 $(SU(3)_c, SU(2)_L)$ gauge group, 
 which are denoted as $Q+{\overline Q}$, are added. 
The non-supersymmetric $SU(5)$ model 
 with ${\bf 10 + \overline{10}}$ 
 (${\bf 15 + \overline{15}}$) Higgs fields
 on an $S^1/(Z_2 \times Z_2')$ 
 orbifold proposed in Ref.\cite{kawamura} 
 has the same field contents at the low 
 energy\footnote{The light (weak scale) masses of 
 scalar fields can not be stable against 
 the radiative corrections because it is 
 the non-supersymmetric theory. 
 The underlying theory which preserves
 light scalar mass from quantum effects is needed
 when we consider the GUT.}. 
Where 
 only leptoquark fields 
 $Q$ and ${\overline Q}$ 
 in ${\bf10} + {\bf\overline{10}}$ 
 (${\bf15} + {\bf\overline{15}}$) Higgs fields 
 have Kaluza-Klein zero modes 
 and survive in the low energy.
In this model the gauge group reduction
 of $SU(5) \rightarrow SU(3)_c \times SU(2)_L \times U(1)_Y$ 
 is realized due to the mass splitting caused by 
 $S^1/(Z_2 \times Z_2')$ orbifolding where 
 the standard gauge bosons have Kaluza-Klein zero modes while
 $X,Y$ gauge bosons do not. 
These two models with additional leptoquarks
 and a Higgs doublet at the low energy unify 
 three gauge coupling constants, but 
 the unification scale is 
 of order $10^{14}$ GeV, which
 disagrees with the present
 proton-decay experiment at Super-Kamiokande.

Thus, 
 in the first model, 
 we introduce additional two ${\bf 24}$ Higgs multiplets 
 in five dimensions, and assign $(Z_2, Z_2')$ parity as 
 $({\bf 8}, {\bf 1}) + ({\bf 1}, {\bf 3}) + ({\bf 1}, {\bf 1})$
 components 
 have Kaluza-Klein zero modes 
 while 
 $({\bf 3}, {\bf 2}) + ({\bf \overline{3}}, {\bf 2})$
 components do not. 
Then the gauge coupling unification is realized
 at the scale of order 
 $10^{17}$ GeV with $O(1)$ TeV gauge invariant 
 mass of ${\bf10} + {\bf\overline{10}}$ 
 $({\bf15} + {\bf\overline{15}})$ and 
 two ${\bf24}$ multiplets. 
The ${\bf24}$ Higgs fields do not have Yukawa couplings 
 with the matter fermions, and  
 the dominant proton-decay mode is 
 $p \rightarrow e^+ \pi^0$ via the exchange of the $X,Y$ gauge 
 bosons which have Kaluza-Klein masses.
However the $X,Y$ gauge 
 bosons are too heavy for the proton decay 
 to be observed at Super-Kamiokande. 
The contributions for the $b \rightarrow s \gamma$, 
 $\mu \rightarrow e \gamma$,  
 electron and muon anomalous magnetic 
 moment, 
 and electron and neutron electric dipole 
 moment from leptoquarks with $O(1)$ TeV masses 
 are negligible. 
Therefore the order one coupling 
 between leptoquarks and quarks/leptons
 are consistent with the current experimental bounds. \\

\noindent
{\bf (2) SUSY flipped $SU(5) \times U(1)$ GUT\cite{HSSU}:} 

The second model we propose is a 
 supersymmetric GUT in five dimensions 
 where the fifth dimensional coordinate is 
 compactified on an $S_1/(Z_2 \times Z_2')$ orbifold. 
In the GUT, the triplet-doublet (TD) 
 splitting problem, 
 how to realize the mass splitting 
 between the triplet and 
 the doublet Higgs particles 
 in the Higgs sector, 
 is one of the most serious problems. 
Recently people show that the 
 five dimensional $SU(5)$ 
 GUT on an $S_1/(Z_2 \times Z_2')$ 
 orbifold can give a natural solution 
 for the TD splitting 
 problem\cite{kawamura2}\cite{HN}\cite{others}. 
In this model, 
 Higgs and gauge fields can propagate in
 five dimensions, while 
 chiral matter fields are localized on the 
 four dimensional wall. 
The TD splitting and the gauge group 
 reduction 
 are realized, 
 since doublet components in ${\bf 5}$ and $\overline{\bf 5}$
 Higgs fields and 
 the standard gauge bosons have Kaluza-Klein zero 
 modes (while other components do not have Kaluza-Klein zero 
 modes) due to the 
 $S_1/(Z_2 \times Z_2')$ orbifolding.

The model we propose is 
 a supersymmetric flipped $SU(5) \times U(1)$ 
 GUT on an $S_1/(Z_2 \times Z_2')$ orbifold, 
 which can realize not only the TD splitting 
 but also the natural fermion mass hierarchies. 
The TD splitting is realized 
 by the $S_1/(Z_2 \times Z_2')$ orbifolding, which 
 also reduces the gauge group as 
 $SU(5) \times U(1) \rightarrow SU(3)_c \times 
 SU(2)_L \times U(1)_Z \times U(1)_X$. 
In addition to three generation chiral matter fields localized on 
 the wall at $y=0$ $(\pi R)$, 
 which are denoted as 
 $({\bf 10}_{1})_i = ( Q_L, 
       \overline{D_R^c}, \overline{N_R^c} )_i$, 
 $({\overline{\bf 5}_{-3})_i} = 
 ( \overline{U_R^c}, L_L  )_i$, and
 $({\bf 1}_{5})_i = ( \overline{E_R^c} )_i$ 
 where index number shows the charge of $U(1)$ and 
 the generation is denoted by $i = 1, 2, 3$, 
 we introduce extra three sets of vector-like  
 matter fields which can propagate in five dimensions. 
The extra 
 matter fields are given by 
 $({\bf 10}_{1})_I+(\overline{\bf 10}_{-1})_I = 
 (Q_L, \overline{D_R^c}, \overline{N_R^c} )_I+
 (\overline{Q_L}, {D_R^c}, {N_R^c} )_I$, 
 $( \overline{\bf 5}_{-3})_I+({\bf 5}_{3})_I = 
 ( \overline{U_R^c}, L_L )_I+({U_R^c}, \overline{L_L} )_I $,
and 
 $({\bf 1}_{5})_I+({\bf 1}_{-5})_I = 
 ( \overline{E_R^c})_I+({E_R^c})_I$,
 where the index shows the extra generation 
 as $I=4,5,6$. 
The suitable $(Z_2, Z_2')$ parity assignment 
 of extra matter fields make 
 $Q_L$, $\overline{Q_L}$, $U_R^c$, $\overline{U_R^c}$,
 $E_R^c$, and $\overline{U_R^c}$ have 
 Kaluza-Klein zero modes and survive bellow 
 the energy scale of compactification. 
On the other hand, other components are heavy enough 
 to be decouple from chiral matter fields. 
The superpotential of the Yukawa sector 
 on the wall $y=0$ $(\pi R)$ is given by
$$
 W_Y = 
 y^d_{ij} H_5 {\bf 10}_i {\bf 10}_j
+ y^u_{ij} H_{\overline{5}} {\bf \overline{5}}_i {\bf 10}_j
+ y^e_{ij} H_5 {\bf \overline{5}}_i {\bf 1}_j 
+ y^{\nu}_{ij} H_{\overline{10}} {\bf 10}_i \phi_j 
+ M_{ij}^\phi \phi_i \phi_j , 
$$
where $\phi_i$ is the gauge singlet matter field 
 localized on the four dimensional wall with 
 the lepton number quantity, 
 and its mass $M_{ij}^\phi$ 
 breaks the lepton number symmetry. 
All Yukawa couplings are 
 assumed to be of $O(1)$ independently 
 of the generation index.
$H$s represent Higgs fields which can 
 propagate in the five dimensions. 
The extra matters can have gauge invariant 
 vector-like mass terms, 
%
$$
 W_{M} = 
   M_{IJ} {\bf 10}_I {\bf \overline{10}}_J
  + M_{IJ}' {\bf 5}_I {\bf \overline{5}}_J
  + M_{IJ}'' {\bf 1}_I {\bf \overline{1}}_J 
  + m_{iI} {\bf 10}_i {\bf \overline{10}}_I
  + m_{iI}' {\bf \overline{5}}_i {\bf 5}_I 
  + m_{iI}'' {\bf 1}_i {\bf \overline{1}}_I 
$$
%
on $y=0$ $( \pi R)$.
{}For simplicity, we take 
 $M_{IJ} = M_{IJ}' = M_{IJ}'' = M_I \delta_{IJ}$  
 and $m_{iJ} = m_{iJ}' = m_{iJ}'' = m_i \delta_{i(J-3)}$. 
We assume the values of $M_{I}$ and $m_{i}$ are smaller
 than the compactification scale of $M_c$, 
 but much larger than the SUSY breaking scale 
 to avoid the blow-up of the gauge coupling constants. 
The Kaluza-Klein zero 
 modes of vector-like matter fields 
 obtain supersymmetric mass terms of $M_I$ and $m_i$.  
The fermion mass hierarchies 
 in the chiral matter fields are generated by 
 integrating out the heavy extra vector-like 
 generations\cite{BB}. 
We consider the case of 
 $\epsilon_1 \simeq M_4/m_1 \ll 1$,
 $\epsilon_2 \simeq M_5/m_2  < 1 $, and 
 $M_6/m_3 \sim 1$, where 
 $\epsilon_i \equiv {M_{i+3} / 
 \sqrt{M_{i+3}^2 + m_i^2}}$. 
Then, 
 the mass hierarchies is generated 
 in the mass matrices of the 
 light eigenstates, $Q_i$, 
 $\overline{U_i}$, and $\overline{E_i}$. 
On the other hand,  
 $\overline{D_i}$, $L_i$, and
 $\overline{N_i}$ do not produce the mass hierarchies 
 since their extra vector-like generations 
 do not have zero modes. 
Then, 
 the light eigenstates mass matrices of 
 up quark sector, down quark sector, 
 and charged lepton sector become 
$$
 m_u^l \simeq \left(
\begin{array}{ccc}
 \epsilon_1^2 & \epsilon_2 \epsilon_1 &  \epsilon_1  \\
 \epsilon_1 \epsilon_2 & \epsilon_2^2  &  \epsilon_2  \\
 \epsilon_1  & \epsilon_2  & 1
\end{array}
\right)  \overline{v}, \;\;
 m_d^l \simeq \left(
\begin{array}{ccc}
 \epsilon_1 & \epsilon_1 &  \epsilon_1  \\
 \epsilon_2 & \epsilon_2  &  \epsilon_2  \\
  1 & 1 & 1 
\end{array}
\right) v , \;\;
 m_e^l \simeq \left(
\begin{array}{ccc}
 \epsilon_1 & \epsilon_2 & 1  \\
 \epsilon_1 & \epsilon_2  & 1 \\
 \epsilon_1 & \epsilon_2 & 1 
\end{array}
\right) v , \;\;
$$
respectively, 
where $v$ and $\overline{v}$ are 
 vacuum expectation values (VEVs) of doublet components
 of $H_5$ and $H_{\overline{5}}$, respectively.
Each element of mass matrices 
 is understood to be multiplied by 
 $O(1)$ coefficient. 
We write the mass matrices 
 that the left-handed fermions are to the left
 and the right-handed fermions are to the right. 
Setting the values of $M_{i+3}$ and $m_i$ as 
 $\epsilon_1 \simeq \lambda^4$ and 
 $\epsilon_2 \simeq \lambda^2$, 
 where $\lambda$ is the Cabbibo angle estimated as $0.2$, 
 we can obtain the suitable fermion mass hierarchies. 
{}Moreover, 
 the small (large) flavor mixings 
 in the quark (lepton) sector are naturally 
 obtained\cite{BB}\cite{BB1}\cite{anarchy}. 
The mass matrix of three light neutrinos is given by 
%
$$
 m_\nu^{(l)} \simeq
\left(
\begin{array}{ccc}
 1 & 1 & 1 \\
 1 & 1 & 1 \\
 1 & 1 & 1 
\end{array}
\right)  {\overline{v}^2 M^\phi \over y^\nu{}^2 v_N^2},
$$
%
where $v_N$ is the VEV of $SU(3)_c \times SU(2)_L$ singlet
 component in $H_{\overline{10}}$ $(H_{10})$, 
 which reduce the gauge group as 
 $SU(3)_c \times 
 SU(2)_L \times U(1)_Z \times 
 U(1)_X \rightarrow SU(3)_c \times 
 SU(2)_L \times U(1)_Y$\footnote{
In order to obtain the large value of $v_N$, 
 we must forbid the mass term of $H_{\overline{10}} H_{10}$.
}. 
The radiative corrections of 
 the large Yukawa coupling $y^{\nu}$
 can induce the VEV, $v_N$. 
We can obtain the suitable mass scale for
 the neutrino oscillation experiments 
 by choosing the values of $M^\phi$ and $y^\nu v_N$. 
The suitable choice of $O(1)$ coefficients 
 in the mass matrix can induce the lepton 
 flavor mixings which are consistent with the neutrino
 oscillation experiments. 
Above fermion mass matrices can give the 
 suitable fermion mass hierarchies of 
 quarks and leptons\cite{BB}\cite{BB1}\cite{anarchy}. 
They also give us the natural explanation 
 why the flavor mixing in the quark sector 
 is small while the flavor mixing in the 
 lepton sector is large.
Since all components have undetermined 
 $O(1)$ coefficients, we have to determine
 the explicit values of $O(1)$ coefficients
 in order to predict the experimentally observable
 quantities. 
As for the proton-decay process, 
 the dimension
 five operator is strongly suppressed by 
 the $U(1)_R$ symmetry, and  the dominant mode is 
 $p \rightarrow e^+ \pi^0$ via the exchange of the $X,Y$ gauge 
 bosons which have Kaluza-Klein masses.


\end{document}